\documentclass[a4paper,10pt,journal]{IEEEtran}
\usepackage{cite}
\usepackage{siunitx}
\DeclareSIUnit \GHz {GHz}
\usepackage{graphicx}
\usepackage{algorithm}
\usepackage{amsfonts}
\usepackage{mathtools }
\usepackage{amsmath}
\usepackage[caption=false,font=footnotesize]{subfig}
\usepackage[hyphens]{url}
\usepackage[noend]{algpseudocode}

\newcommand{\FF}[1]{{\mathbb{F}}}

\usepackage[inkscapeformat=pdf]{svg}
\usepackage{tikz}
\usetikzlibrary{backgrounds}
\usetikzlibrary{arrows,automata}

\usepackage{enumerate}

\newtheorem{remark}{Remark}[section]

\newcolumntype{P}[1]{>{\centering\arraybackslash}p{#1}}
\newcolumntype{M}[1]{>{\centering\arraybackslash}m{#1}}
\begin{document}
\title{{\LARGE On Optimization of Next-Generation Microservice-Based Core Networks}}

\author{Andrea Tassi, Daniel Warren, Yue Wang, Deval Bhamare, and Rasoul Behravesh
\thanks{
A. Tassi, Y. Wang, D. Bhamare and R. Behravesh are with Advanced Network Research, and D. Warren is with Communications Research, \mbox{Samsung} R\&D Institute UK (SRUK), UK (e-mail: {\tt \{a.tassi, dan.warren, yue2.wang\}@samsung.com}).
}}

\maketitle

\begin{abstract}
Next-generation mobile core networks are required to be scalable and capable of efficiently utilizing heterogeneous bare metal resources that may include edge servers. To this end, microservice-based solutions where control plane procedures are deconstructed in their fundamental building blocks are gaining momentum. This letter proposes an optimization framework delivering the partitioning and mapping of large-scale microservice graphs onto heterogeneous bare metal deployments while minimizing the total network traffic among servers. An efficient heuristic strategy for solving the optimization problem is also provided. Simulation results show that, with the proposed framework, a microservice-based core can consistently support the requested load in heterogeneous bare metal deployments even when alternative architecture fails. Besides, our framework ensures an overall reduction in the control plane-related network traffic if compared to current core architectures.
\end{abstract}


\vspace{-2mm}\section{Introduction}\label{sec:intro}\vspace{-1mm}
The standard documents for the Third Generation Partnership Project's (3GPP's) Release 15 core network (known as ``fifth generation core network’’ or ``5GC’’) define a clear separation between the Control Plane (CP) and User Plane (UP) procedures~\cite{3gpp.23.502}. CP handles the control signaling of the mobile network; UP carries the actual user traffic. 

This letter focuses on CP procedures within the 5GC Service-Based Architecture (SBA)~\cite{3gpp.23.502}. UP and data plane optimization is out of the scope of this paper. Currently, CP procedures are delivered through a plurality of services, which are, in turn, clustered into Network Functions (NFs), hereafter referred to as ``\emph{NF-based}'' architecture. This letter describes and optimizes the bare metal execution of a microservice-based (\emph{MS-based}) core network where each of 3GPP's NF is deconstructed into its fundamental functional services. In so doing, the basic tasks carried out by each functional building block are delivered by a single MS~\cite{MS0}. 
Beyond the context of 5GC, several studies have focussed on the issue of allocating chains of NF instances (also known as Service Function Chains, SFCs) onto an available set of computation resources (e.g., bare metal servers, virtual machine instances, etc.). In particular, Y. Yue \emph{et al.}~\cite{R1} and B. Ren \emph{et al.}~\cite{R2} formulate bare metal resource mapping frameworks for SFCs accounting for both the network and processing overhead determined by the NF instances forming each SFC. The general problem of SFC bare metal placement has also been extended by B. N\'emeth \emph{et al.}~\cite{R3} to account for issues pertaining to the radio access network (such as network coverage and reliability). J. Sun \emph{et al.}~\cite{R4} further extended the modeling framework of the SFC placement problem by establishing a link between network traffic processing carried out by an NF instance and its CPU footprint.
In the context of 5GC, E. Goshi~\emph{et al.}~\cite{goshi2022pp5gs} propose a \emph{procedure-based} core network where inter-NF communications are minimized by substituting 3GPP's defined NFs with procedural-based NFs, where a single monolithic NF delivers one CP procedure.

Unlike~\cite{goshi2022pp5gs}, this letter focuses on an MS-based architecture~\cite{MS0} suitable for implementing all the CP procedures a 5GC core network (and beyond) carries out. Current works~\cite{MS0,MS1} consider clusters of MS instances potentially deployed at the edge to deliver a high-level use case (e.g., Virtual Reality, Edge Processing) but leaving the core network essentially unaltered. This letter advocates for a paradigm shift where an MS-based core network makes it possible to utilize heterogeneous bare metal deployments efficiently.
Even though MS-based systems simplify the adoption of a high level of software development hygiene, the topic of Security of distributed systems is beyond the scope of this paper.
We propose an optimization framework to model MS-based CP procedures by describing logical interactions among MSs and their computational and network footprints. We model the resource allocation problem of an MS-based core network as the problem of mapping multiple MS instances onto a given set of servers without violating either the maximum server computational or server network capacities. 
Unlike~\cite{R1,R2,R3} where the network traffic generated by each NF instance (MS instances, in our case) is pre-determined, our optimization framework captures the difference in throughput that can be achieved between a pair of MS instances mapped onto the same or different servers. If the latter, the overhead caused by the task of handling the transmission of a stream of network packets through a physical network interface controller notably reduces the inter-MS instance communication throughput, if compared to the case where instances are co-located in the same server~\cite{DPDK}. We observe that a modeling effort in this direction has been attempted in~\cite{R4}, where the computation footprint associated with an NF instance is a function of network traffic provided as an input to the NF instance itself; the performance model considered in~\cite{R4} does not differentiate between network traffic exchanged between co-located and non-co-located NF instances. Overall, the modeling effort in~\cite{R1,R2,R4} is steered towards the issue of mapping multiple short SFCs (consisting of ten elements or less) onto the available servers. At the same time, the formulation of our optimization framework focuses on how to partition large MS graphs (formed by hundreds of MS instances), delivering multiple procedures and mapping them on the available servers. The proposed optimization framework will favor the packing of MS instances of the same CP procedure to minimize the network traffic among servers. An efficient heuristic strategy for calculating feasible solutions to the problem above is also provided.

The rest of the paper is structured as follows. Section~\ref{sec:systemModel} presents the considered system model. The proposed resource allocation problem and the procedure to calculate its heuristic solution are described in Sections~\ref{sec:opt} and~\ref{sec:heuristic}, respectively. A performance evaluation of the proposed approach is given in Section~\ref{sec:evaluation}. Finally, in Section~\ref{sec:conclusion}, we draw our conclusions.

\vspace{-2mm}\section{System Model}\label{sec:systemModel}\vspace{-1mm}
We define the bare metal resources as tuple $(\mathcal{S},\mathbf{E},\mathcal{R},\mathbf{N})$. Key notations are summarized in Table~\ref{tab.1}. Set $\mathcal{S} = [|\mathcal{S}|]$ consists of the available servers\footnote{For positive integer $K$, $[K]$ denotes the set $\{1, 2, \ldots, K\}$.}. We regard $\mathbf{E}$ as the $|\mathcal{S}| \times |\mathcal{S}|$ \emph{network adjacency matrix} where $(i,j)$-th element $e_{i,j}$ is equal to $1$ if a network link is present between servers $i$ and $j$, and $0$, otherwise. If $i$ is equal to $j$, $e_{i,j}$ is assumed to be equal to $1$ since intra-server communication should always be possible. If servers $a$ and $b$ (for $a,b \in S$) are distinct and $e_{a,b} = 1$, then $e_{b,a} = 1$ also. The network interconnecting the available servers may be partitioned into a number of sub-networks, where a server may communicate with other servers pertaining to the same or a different sub-network. Servers do not take part in any intra-/extra-sub-network network traffic forwarding tasks -- such tasks are considered to be performed by network switches and routers. For the sake of compactness in our notation, we say that a network link exists between servers $i$ and $j$ (for $i \neq j$ and $i,j \in S$) if any of the following facts hold true: (i) $i$ and $j$ belong to the same sub-network, and $i$ communicates with $j$ via a number of switches (or none) or (ii) $i$ and $j$ belong to two different sub-networks, and $i$ communicates with $j$ via a number of switches (or none) and one or more routers without their traffic having to go through a third sub-network.

The server resource capacities are listed in set $\mathcal{R} = \{(r^{(\textrm{CPU})}_i, r^{(\textsc{Mem})}_i), \forall i \in S\}$, where tuple $(r^{(\textrm{CPU})}_i, r^{(\textsc{Mem})}_i)$ consists of the maximum CPU capacity that can be utilized and the maximum amount of memory that can be allocated in server $i$, respectively. CPU capacities are expressed as the sum of the single capacities (ranging between $0$ and $1$) that can be allocated on each CPU core, which is then normalized with respect to the maximum CPU capacity supported by all the servers. Similarly, memory capacities are also normalized by the maximum amount of memory that any server can allocate.  
We regard $\mathbf{N}$ as the $|\mathcal{S}| \times |\mathcal{S}|$ \emph{link capacity matrix} where element $n_{i,j}$ signifies the capacity associated with the link connecting server $i$ to server $j$, measured in terms of protocol data units (PDUs) per second. If there is no link between a pair of servers ($e_{i,j}$ is equal to 0), the $n_{i,j}$ is also equal to $0$. Capacities of links originating and directed to the same server are set equal to infinity ($n_{a,a} = \infty$).

\vspace{0.0cm}Only those core MSs that receive inputs from the Radio Access Network (RAN) and/or other MSs and produce outputs intended for the RAN and/or other MSs are considered in this letter. RAN-specific operations are not considered in the proposed resource allocation model.  Each input request determines (i) the nature of the computation to be carried out by the MS and (ii) towards which other MS or RAN component the output is expected to be directed.

\vspace{0.0cm}We define the CP procedure $t$ by tuple $P^{(t)} = (\mathcal{M}^{(t)},\mathbf{V}^{(t)},\mathcal{F}^{(t)},A^{(t)},\mathcal{C}^{(t)},\mathcal{\Hat{L}}^{(t)})$, where $\mathcal{M}^{(t)} = [|\mathcal{M}^{(t)}|]$ is the list of the MSs pertaining to the CP procedure. The logical interaction between the MSs is defined by a $|\mathcal{M}^{(t)}| \times |\mathcal{M}^{(t)}|$ \emph{MS adjacency matrix} matrix $\mathbf{V}^{(t)}$ where its $(i,j)$-th element $v_{i,j}^{(t)}$ is equal to $1$ if MS $i$ directly communicates with MS $j$, and $0$, otherwise. The resource footprint associated with each MS, while it processes an input request, is defined by set $\mathcal{F}^{(t)} = \{(f^{(t,\textrm{CPU})}_i, f^{(t,\textsc{Mem})}_i), \forall i \in \mathcal{M}^{(t)}\}$, where tuple $(f^{(\textrm{t,CPU})}_i, f^{(t,\textsc{Mem})}_i)$ defines the CPU and memory footprint of MS $i$, respectively. The footprints of MSs not processing any input requests are negligible and, thus, are not considered. MS footprints are normalized by the aforementioned factor for the server resources.
Term $a^{(t)}_{i,j}$ defines the $(i,j)$-th element of a $|\mathcal{M}^{(t)}| \times |\mathcal{M}^{(t)}|$ matrix $\mathbf{A}^{(t)}$, where $a^{(t)}_{i,j}$ is the time MS $i$ needs to process an input request and generate an output intended for MS $j$ -- $\mathbf{A}^{(t)}$ will be referred to as \emph{base MS processing time matrix}. In particular, if MS $i$ directly communicates to MS $j$ ($v_{i,j}^{(t)} = 1$),  $a^{(t)}_{i,j}$ is set to an operationally-realistic value greater than $\SI{0}{\second}$. If $v_{i,j}^{(t)} = 0$, $a^{(t)}_{i,j}$ is set equal to the largest positive machine number. If an instance of MS $i$ is generating outputs for another instance located on a different server, being set $\mathcal{C}^{(t)} = \{c^{(t)}_i, \forall i \in \mathcal{M}^{(t)}\}$, the time MS $i$ needs to process each input request and generate output is increased of $c^{(t)}_i$ seconds. Equivalently, $\mathcal{C}^{(t)}$ consists of the extra processing time that each MS instance requires (on top of the base MS processing time) to generate an output directed to a remote instance.
Set $\mathcal{L}^{(t)} = \{\ell^{(t)}_i, \forall i \in \mathcal{M}^{(t)}\}$ is the maximum number of concurrent requests (originating from gNBs or other MSs) that each MS supports before experiencing performance degradation. The aforementioned set consists of the maximum request load supported by each MS. Set $\mathcal{P} = \{P^{(1)}, P^{(2)}, \ldots, P^{(|\mathcal{P}|)}\}$ comprises of the CP procedures supported by the system. This system model can be used to describe a system as per in the following remark.
\begin{remark}\label{rem.implem}
With regards to the 3GPP's stage 3 definition of any CP procedure~\cite{3gpp.23.502}, a CP procedure consists of a number of MSs equal to the number of its \emph{distinct} service-operation pairs (e.g.,\cite[Table 5.2.2.1-1]{3gpp.23.502}); one MS per service-operation pair. Since procedures are standardized as sequences of service-operation pairs~\cite{3gpp.23.502}, it is immediate to establish the interactions among MSs. In a Docker-powered Kubernetes environment\cite{MS0}, each MS instance is run in a container. In each server, all the containers running MS instances pertaining to the same CP procedure are clustered in a pod.
\end{remark}

\begin{table}[t]
\vspace{-2.5mm}\scriptsize
\renewcommand{\arraystretch}{1}
\caption{Key notation used throughout the paper.}
\centering
\vspace{-1mm}\begin{tabular}{|m{2cm}|m{5.9cm}|}
\hline
\textbf{Notation}               & \textbf{Description}\\
\hline
\hline $\mathcal{S}$                      & Available servers \\
\hline $\mathbf{E}$, $e_{i,j}$           & Network adjacency matrix and its $(i,j)$-th element \\
\hline $\mathcal{R}$,$\;$ $(r^{(\textrm{CPU})}_i, r^{(\textsc{Mem})}_i)$         & Server resource capacities, and tuple expressing the CPU, memory capacity that can be allocated in server $i$ \\
\hline $\mathbf{N}$, $n_{i,j}$           & Link capacity matrix and its $(i,j)$-th element\\
\hline $\mathcal{P}$, $P^{(t)}$           & Set of supported CP procedure and $t$-th procedure \\
\hline $\mathcal{M}^{(t)}$                & List of the MSs pertaining to CP procedure $t$\\
\hline $\mathbf{V}^{(t)}$, $v_{i,j}^{(t)}$   & MS adjacency matrix and its $(i,j)$-th element pertaining to CP procedure $t$\\
\hline $\mathcal{F}^{(t)}$, $(f^{(\textrm{t,CPU})}_i, f^{(t,\textsc{Mem})}_i)$                & MS footprints, and tuple expressing the CPU and memory footprint of MS $i$ of CP procedure $t$ \\
\hline $\mathbf{A}^{(t)}$, $a^{(t)}_{i,j}$                & Base MS processing time matrix and its $(i,j)$-th element associated with CP procedure $t$\\
\hline $\mathcal{C}^{(t)}$, $c^{(t)}_i$             & Set of the extra processing time that each MS instance requires to generate an output directed to a remote instance and its $i$-th element associated with CP procedure $t$ \\
\hline $\mathcal{\Hat{L}}^{(t)}$, $\ell^{(t)}_i$          & Set of the maximum request load of each MS, and its $i$-th element associated with CP procedure $t$ \\
\hline $\tau^{(t)}_i$          & Number of instances associated with MS $i$ pertaining to CP procedure $t$ \\
\hline
\end{tabular}\vspace{-2mm}
\label{tab.1}
\end{table}

\vspace{-0.5cm}\section{Proposed Optimization Framework}\label{sec:opt}
Let $\hat{U}^{(t)}$ be the number of concurrent user requests for CP procedure $t$, for $t \in [|\mathcal{P}|]$. We set the number $\tau^{(t)}_i$ of instances associated with MS $i$ (for $i \in \mathcal{M}^{(t)}$) pertaining to CP procedure $t$ as $\tau^{(t)}_i = \left\lceil\hat{U}^{(t)}/\ell^{(t)}_i\right\rceil$.

In the remainder of this section, we will define the Core Optimization (CO) problem. Given a number of MS instances across all the considered CP procedures $\{\tau^{(t)}_i,  \forall t \in [|\mathcal{P}|], \forall i \in \mathcal{M}^{(t)}\}$, CO can be solved by the MS instance allocation that maps all the MS instances onto the available servers such that the overall flow that may circulate across all the network links is minimized. Crucially, an MS instance-server mapping is regarded as feasible iff the following facts hold true: (i) each MS instance is mapped on only one server, (ii) MS instances expected to change processing results need to be mapped onto the same server or onto servers that can directly communicate among each other, (iii) the server resource and link capacities are not violated. The CO problem is defined as follows.

We define the binary optimization variable $x^{(t,i)}_{r,s}$ (for $t \in [|\mathcal{P}|]$, $i \in \mathcal{M}^{(t)}$, $r \in [\tau^{(t)}_i]$ and $s \in \mathcal{S}$) to be equal to $1$ if instance $r$ of MS $i$, pertaining to CP procedure $t$, is mapped onto server $s$, $0$ otherwise. The following constraint ensures that each replica of any MS is mapped onto only one server.
\begin{align}
  \sum_{s = 1}^{|\mathcal{S}|} x^{(t,i)}_{r,s} = 1, \quad \forall t \in [|\mathcal{P}|], \forall i \in \mathcal{M}^{(t)}, \forall r \in [\tau^{(t)}_i]. \label{COM.c1}
\end{align}

For any given pair of MSs $(i,j)$, pertaining to a CP procedure $t$, such that MS $i$ is required to send its outputs to MS $j$ (namely, $v^{(t)}_{i,j} = 1$), there might be more than one instance of MS $i$ and/or MS $j$ ($\tau^{(t)}_i \geq 1$ and/or $\tau^{(t)}_j \geq 1$). In this case, for simplicity, we impose that all the instances of MS $i$ and $j$ are mapped onto servers that can directly communicate with each other (including instances mapped onto the same server). As such, if instance $r$ of MS $i$ is mapped onto server $s$ ($x^{(t,i)}_{r,s} = 1$) and instance $y$ of MS $j$ is mapped onto server $o$ ($x^{(t,j)}_{y,o} = 1$) then server $s$ must be able to communicate with server $o$ ($e_{s,o} = 1$). The constraint can be equivalently expressed as follows~\cite{math10020283}:
\begin{align}
  \hspace{-0.5cm}v^{(t)}_{i,j}x^{(t,i)}_{r,s}x^{(t,j)}_{y,o} (e_{s,o} - 1&) = 0, \quad \forall t \in [|\mathcal{P}|], \forall i,j \in M^{(t)},\notag\\
  &\forall s,o\in \mathcal{S}, \forall r\in [\tau_i^{(t)}], \forall y\in [\tau_j^{(t)}].\hspace{-0.2cm} \label{COM.c2}
\end{align}

The following constraints ensure the overall CPU and memory footprints of all the MS instances allocated onto a server are always smaller than or equal to the resource capacity of the server.
\begin{align}
  \sum_{\mathclap{\substack{t \in [|\mathcal{P}|], \; i \in [|\mathcal{M}^{(t)}|] \\ r \in [\tau^{(t)}_i] }}} x^{(t,i)}_{r,s} f^{(t,k)}_i \leq r^{(k)}_s, \forall s \in \mathcal{S}, \forall k \in \{\textrm{CPU}, \textsc{Mem}\}. \hspace{-0.4cm} \label{COM.c3}
\end{align}
The constraint above ensures that the overall footprint of all the MS instances mapped onto a server will not exceed the server capacity even when all the MS instances process an input request simultaneously. 

The dynamic optimization of the incoming request load experienced by each MS instance is beyond the scope of this letter. However, in formulating the CO problem, we assumed the presence of a load balancer regulating the flows among all the MS instances. As such, assuming MS $i$ is expected to send its outputs to MS $j$ ($v^{(t)}_{i,j}$), the output flow of an instance of MS $i$ directed to one or more instances of MS $j$ is uniformly distributed across the $\tau^{(t)}_j$ instances of MS $j$. More formally, the flow produced by one instance of MS $i$ and directed to one instance of MS $j$ (in case $i$ continuously processes input requests determining outputs intended for $j$) is equal to $1/(\tau^{(t)}_j a^{(t)}_{i,j})$ or $1/(\tau^{(t)}_j(a^{(t)}_{i,j} + c^{(t)}_{i,j}))$ if the instances are co-located in the same server or not, respectively. Overall, if $\xi_{a,b}$ signifies the total flow associated with the network link connecting server $a$ to server $b$, for $a,b \in [|\mathcal{S}|]$. The following constraint ensures that $\xi_{a,b}$ does not exceed the capacity of the network link.
\begin{align}
  \hspace{0.5cm}\overbrace{\sum_{\mathclap{\substack{t \in [|\mathcal{P}|] \\ i,j \in [|\mathcal{M}^{(t)}|]\\\ell \in [\tau^{(t)}_i], q \in [\tau^{(t)}_j] }}} \quad \frac{v^{(t)}_{i,j} x^{(t,i)}_{\ell,a} x^{(t,j)}_{q,b}}{\tau^{(t)}_j(a^{(t)}_{i,j} + \delta_{(a \neq b)} c^{(t)}_{i,j})}}^{\xi_{a,b}} &\leq n_{a,b}, \quad \forall a,b \in \mathcal{S}, \label{COM.c4}
\end{align}
where the term $\delta_{(H)}$ is equal to $1$ if statement $H$ is true, $0$ otherwise. We observe that the term $v^{(t)}_{i,j} x^{(t,i)}_{\ell,a} x^{(t,j)}_{q,b}$ is equal to $1$ iff an instance of MS $i$ provides input to an instances of MS $j$, and instance $\ell$ of MS $i$ and instance $q$ of MS $j$ are mapped onto servers $a$ and $b$, respectively. Constraint~\eqref{COM.c4} ensures that the capacity of any network link is not exceeded even when all the MS instances generate outputs at their maximum rate.

We define the Core Optimization (CO) problem as follows.
\vspace{-5mm}\begin{align}
  \text{CO} &  \quad  \min_{\mathbf{x}} \,\,  \displaystyle\overbrace{\sum_{\mathclap{\substack{a,b \in [|\mathcal{S}|]}}} \delta_{(a \neq b)}\xi_{a,b}}^{\Psi(\mathbf{x})}, \label{COM.of1}\\
\text{s.t.} &  \quad\eqref{COM.c1}, \eqref{COM.c2}, \eqref{COM.c3} \text{ and } \eqref{COM.c4}. \label{COM.c0}
\end{align}
where $\mathbf{x} = \{x^{(t,i)}_{r,s}, \forall t \in [|\mathcal{P}|], \forall i \in \mathcal{M}^{(t)}, \forall r \in [\tau_i], \forall s \in \mathcal{S}\}$. In particular, $\Psi(\mathbf{x})$ represents the overall flow that may circulate across all the network links but disregards the flow among MS instances running on the same server.
We observe that an optimum solution to the CO problem will minimize the total network traffic between servers by favoring the packing of MSs instances pertaining to the same CP procedure onto the same server.

\vspace*{-0.5cm}\section{Proposed MS Mapping Procedure}\label{sec:heuristic}
\begin{figure}[t]
\vspace{-5.9mm}
\begin{algorithm}[H]
\floatname{algorithm}{Procedure}
 \begin{algorithmic}[1]
\caption{MS Mapping Procedure.}\label{alg:MMP}
\begin{scriptsize}
\Procedure{$\textsc{MM}(t)$}{}
    \State $x_{r,s}^{(t,i)} \gets 0$, $\forall i \in \mathcal{M}^{(t)}$, $\forall r \in [\tau_i^{(t)}]$, $\forall s \in S$ \label{proc1.initA}
    \State $\omega_i \gets 0$, $\forall i \in \mathcal{M}^{(t)}$ \label{proc1.initB}
    \State $\lambda \gets \max_{i \in \mathcal{M}^{(t)}}\{\tau_i^{(t)}\}$ \label{proc1.initC}
    
    \While{ $\lambda > 0$} \label{proc1.OuterLoopStart}
     \State $\overline{s} \gets 0$, $\overline{f} \gets -1$, $\overline{c} \gets 0$
     \State $\mathbf{K} \gets [(\mathcal{M}^{(t)}, \mathbf{V}^{(t)}, \mathcal{F}^{(t)}, \mathbf{A}^{(t)},\mathcal{C}^{(t)},0)]$ \label{proc1.initK}
     \While{$K$ is not empty} \label{proc1.InnerLoopStart}
        \State $(\mathcal{M}, \mathbf{V}, \mathcal{F}, \mathbf{A}, \mathcal{C},\Delta^{(\textsc{Out})}) \gets \text{\textsc{pop}}(K)$
        \State $\Delta^{(\mathrm{CPU})} \gets \sum_{i \in M} f^{(\textsc{CPU})}_i \delta_{(\omega_i < \tau_i)}$,
        \State $\Delta^{(\mathrm{Mem})} \gets \sum_{i \in M} f^{(\textsc{Mem})}_i \delta_{(\omega_i < \tau_i)}$
        
        \State  $s$ $\gets$ $\text{BFF}(\overline{s}, \overline{f}, \Delta^{(\textsc{Out})}, \Delta^{(\mathrm{CPU})}, \Delta^{(\mathrm{Mem})})$ 
            
        \If{$s > 0$}
            \For{$i \in M$ such that $\omega_i < \tau_i$}
                    \State $\omega_i \gets \omega_i + 1$
                    \State $x_{\omega_i,s}^{(t,i)} \gets 1$
            \EndFor
            \State $\overline{s} \gets s$, $\overline{f} \gets \Delta^{(\textsc{Out})}$
        \Else
            \If{$|M| > 1$}
                \State $P^\prime, P^{\prime\prime} \gets \textsc{GP}((\mathcal{M}, \mathbf{V}, \mathbf{A}, \mathcal{C}))$
                \State $\text{\textsc{head}}(K) \gets P^\prime, P^{\prime\prime}$
            \EndIf
            \State \textbf{else} \textbf{return} No solution can be found.
        \EndIf 
        \EndWhile\label{proc1.InnerLoopStop}
        \If{Constraints~\eqref{COM.c4} are violated}
        \State \textbf{return} No solution can be found.
        \EndIf
        \State $\lambda \gets \lambda - 1$
    \EndWhile \label{proc1.OuterLoopStop}
    
    \State \textbf{return} $\{ x_{r,s}^{(t,i)} | i \in \mathcal{M}^{(t)}, r \in [\tau_i^{(t)}], s \in \mathcal{S}\}$.
    \EndProcedure
\end{scriptsize}
\end{algorithmic}
\end{algorithm}
\vspace{-5mm}
\end{figure}

In this section, we propose the MS Mapping (MM)
procedure as an efficient way to solve the CO problem heuristically.  Procedure~\ref{alg:MMP} defines the MM procedure ($\textsc{MM}(t)$) for CP procedure $t$, for $t \in [|\mathcal{P}|]$. To calculate a feasible solution to the CO problem, the MM procedure has to be executed once for each procedure in $\mathcal{P}$. 

Intuitively, Procedure~\ref{alg:MMP} leverages upon a divide-et-impera approach where a \emph{mapping task} is iteratively executed. Each time the mapping task runs, an MS graph associated with CP procedure $t$ only considering \emph{one} MS instance (for all the MS in $\mathcal{M}^{(t)}$) at a time is allocated onto the available servers -- the aforementioned graph will be referred to as \emph{one-MS instance graph}. Depending on the number of concurrent user requests, the number of instances associated with each MS is expected to be greater than one and likely to vary. For these reasons, the mapping task is repeated $\lambda = \max_{i \in M^{(t)}}\{\tau_i^{(t)}\}$ times with the caveat that, in the end, exactly $\tau_i^{(t)}$ instances are mapped onto the available servers, for each MS $i \in \mathcal{M}^{(t)}$. For each one-MS graph to be allocated, the mapping task progressively partitions the graph by establishing the cut associated with the minimum flow, which leads to two sub-graphs. Suppose there are not enough computational resources or network flow capacity available to accommodate all the MS instances of a sub-graph in the same server. In that case, the sub-graph is further partitioned in a similar fashion. The partition process is repeated until all the MS instances are mapped onto the available servers.

More formally, for a given CP procedure $t \in [|\mathcal{P}|]$, the term $\omega_i$ in Procedure~\ref{alg:MMP} counts the number of MS instances of MS $i$ (for $i \in \mathcal{M}^{(t)}$) that have been mapped onto any given server, which at the beginning of the procedure are initialized to $0$ along with the optimization variables (lines~2-3). 
The outer while-loop of $\textsc{MM}(t)$ (lines~5-25) carries out the mapping task, which operates as follows. List $\mathbf{K}$ is initialized (line~7) with (part of) the terms in tuple $\mathcal{P}^{(t)}$, and with the outgoing flow $\Delta^{(\textsc{Out})}$ component (defined further down for the sake of clarity) set to $0$. When the mapping task is executed, it calculates the overall amount of CPU ($\Delta^{(\mathrm{CPU})}$) and memory ($\Delta^{(\mathrm{Mem})}$) footprints associated with the element taken from the head of $\mathbf{K}$ (function $\textsc{pop}(\mathbf{K})$) -- in this case, the entire one-MS instance graph. Then, a Best-Fit-First (BFF) procedure~\cite{BFF} (not included in the letter) attempts to select a server $s$ (line~12) that can accommodate all the MS instances contained in $\mathcal{M}$. If $\textsc{BFF}$ returns a valid server selection, optimization variables pertaining to CP procedure $t$ and $\{\omega_i, \forall i \in \mathcal{M}\}$ are updated (lines 14-16). Otherwise, the one-MS instance graph is partitioned into two sub-graphs via the Graph Partitioning (GP) procedure\footnote{The GP procedure partitions $\mathcal{M}$ into two disjoint sets $\mathcal{M} = \{\mathcal{M}_1, \mathcal{M}_2\}$ while minimizing $\sum_{i \in \mathcal{M}_1, j \in \mathcal{M}_2, v_{i,j} = 1} 1 / (\tau_j(a_{i,j} + c_i))$.} (not included). Each sub-graph will define a tuple ($P^\prime$ and $P^{\prime\prime}$) with the same structure as the tuple considered at line 7, but this time, the outgoing flow term in $P^\prime$ and $P^{\prime\prime}$ is set equal to the overall flow originating from the MS instances associated with one sub-graph and directed to instances that are not part of the considered sub-graph. Tuples $P^\prime$ and $P^{\prime\prime}$ are inserted at the head of the list $\mathbf{K}$, and the internal while-loop of the mapping task (lines~8-22) iterates again until list $\mathbf{K}$ is empty, i.e., until all the MS instances in the one-MS instance graph have been mapped onto the available servers.
Except for the first iteration of the while-loop of the mapping task (lines~8-22), the BFF procedure ensures that the capacity of the link connecting server $\overline{s}$ (identified during a previous iteration of the inner while-loop) to $s$ (identified in the current iteration) is not violated when an additional flow equal to $\overline{f}$ is allocated on the link. The behaviors of the mapping task are further described in the example in Procedure~\ref{alg:MMP} where the while-loop at lines~8-22 iterates three times (and the one-MS instance graph is progressively partitioned into three sub-graphs) before the BFF procedure can map all the four MS instances.

\AddToHookNext{shipout/background}{
  \begin{tikzpicture}[remember picture,overlay]
  \draw[fill=gray, fill opacity=0.2, rounded corners, draw=none] (10.16,-10) rectangle (2.5,-3.95);
  \node[draw=none,align=left, rotate=90] at (2.3,-6.8) {\emph{{\small Mapping Task}}};
  \node[xshift=-1.45cm,yshift=8.1cm] at (current page.center) {
    \tikz \node [scale=0.365, inner sep=0] {

\tikzset{every picture/.style={line width=0.75pt}} 

\begin{tikzpicture}[x=0.75pt,y=0.75pt,yscale=-1,xscale=1]

\draw  [fill={rgb, 255:red, 57; green, 158; blue, 23 }  ,fill opacity=1 ][line width=2.25]  (30,64.5) .. controls (30,52.63) and (39.63,43) .. (51.5,43) .. controls (63.37,43) and (73,52.63) .. (73,64.5) .. controls (73,76.37) and (63.37,86) .. (51.5,86) .. controls (39.63,86) and (30,76.37) .. (30,64.5) -- cycle ;
\draw  [fill={rgb, 255:red, 57; green, 158; blue, 23 }  ,fill opacity=1 ][line width=2.25]  (117,64.5) .. controls (117,52.63) and (126.63,43) .. (138.5,43) .. controls (150.37,43) and (160,52.63) .. (160,64.5) .. controls (160,76.37) and (150.37,86) .. (138.5,86) .. controls (126.63,86) and (117,76.37) .. (117,64.5) -- cycle ;
\draw  [fill={rgb, 255:red, 57; green, 158; blue, 23 }  ,fill opacity=1 ][line width=2.25]  (205,64.5) .. controls (205,52.63) and (214.63,43) .. (226.5,43) .. controls (238.37,43) and (248,52.63) .. (248,64.5) .. controls (248,76.37) and (238.37,86) .. (226.5,86) .. controls (214.63,86) and (205,76.37) .. (205,64.5) -- cycle ;
\draw  [fill={rgb, 255:red, 57; green, 158; blue, 23 }  ,fill opacity=1 ][line width=2.25]  (117,154.5) .. controls (117,142.63) and (126.63,133) .. (138.5,133) .. controls (150.37,133) and (160,142.63) .. (160,154.5) .. controls (160,166.37) and (150.37,176) .. (138.5,176) .. controls (126.63,176) and (117,166.37) .. (117,154.5) -- cycle ;
\draw  [fill={rgb, 255:red, 57; green, 158; blue, 23 }  ,fill opacity=1 ][line width=2.25]  (205,154.5) .. controls (205,142.63) and (214.63,133) .. (226.5,133) .. controls (238.37,133) and (248,142.63) .. (248,154.5) .. controls (248,166.37) and (238.37,176) .. (226.5,176) .. controls (214.63,176) and (205,166.37) .. (205,154.5) -- cycle ;
\draw [line width=0.75]    (74,64.5) -- (116,64.5) ;
\draw [shift={(118,64.5)}, rotate = 180] [color={rgb, 255:red, 0; green, 0; blue, 0 }  ][line width=0.75]    (17.49,-5.26) .. controls (11.12,-2.23) and (5.29,-0.48) .. (0,0) .. controls (5.29,0.48) and (11.12,2.23) .. (17.49,5.26)   ;
\draw [line width=0.75]    (161,154.5) -- (204,154.5) ;
\draw [shift={(206,154.5)}, rotate = 180] [color={rgb, 255:red, 0; green, 0; blue, 0 }  ][line width=0.75]    (17.49,-5.26) .. controls (11.12,-2.23) and (5.29,-0.48) .. (0,0) .. controls (5.29,0.48) and (11.12,2.23) .. (17.49,5.26)   ;
\draw [line width=0.75]    (52.5,86) .. controls (53,146.2) and (46.56,155.9) .. (116.93,154.52) ;
\draw [shift={(118,154.5)}, rotate = 178.8] [color={rgb, 255:red, 0; green, 0; blue, 0 }  ][line width=0.75]    (17.49,-5.26) .. controls (11.12,-2.23) and (5.29,-0.48) .. (0,0) .. controls (5.29,0.48) and (11.12,2.23) .. (17.49,5.26)   ;
\draw [line width=0.75]    (141.33,86.82) .. controls (170.65,99.76) and (174.75,93.71) .. (211.5,79.5) ;
\draw [shift={(139.5,86)}, rotate = 24.3] [color={rgb, 255:red, 0; green, 0; blue, 0 }  ][line width=0.75]    (17.49,-5.26) .. controls (11.12,-2.23) and (5.29,-0.48) .. (0,0) .. controls (5.29,0.48) and (11.12,2.23) .. (17.49,5.26)   ;
\draw [line width=0.75]    (139.5,43) .. controls (160.57,29.77) and (181.64,36.71) .. (207.89,49.23) ;
\draw [shift={(209.5,50)}, rotate = 205.71] [color={rgb, 255:red, 0; green, 0; blue, 0 }  ][line width=0.75]    (17.49,-5.26) .. controls (11.12,-2.23) and (5.29,-0.48) .. (0,0) .. controls (5.29,0.48) and (11.12,2.23) .. (17.49,5.26)   ;
\draw [line width=0.75]    (239.5,137.5) .. controls (249.79,113.49) and (248.56,111.09) .. (238.15,86.53) ;
\draw [shift={(237.5,85)}, rotate = 67.07] [color={rgb, 255:red, 0; green, 0; blue, 0 }  ][line width=0.75]    (17.49,-5.26) .. controls (11.12,-2.23) and (5.29,-0.48) .. (0,0) .. controls (5.29,0.48) and (11.12,2.23) .. (17.49,5.26)   ;
\draw  [fill={rgb, 255:red, 57; green, 158; blue, 23 }  ,fill opacity=1 ][line width=2.25]  (30.5,228) .. controls (30.5,216.13) and (40.13,206.5) .. (52,206.5) .. controls (63.87,206.5) and (73.5,216.13) .. (73.5,228) .. controls (73.5,239.87) and (63.87,249.5) .. (52,249.5) .. controls (40.13,249.5) and (30.5,239.87) .. (30.5,228) -- cycle ;
\draw  [fill={rgb, 255:red, 57; green, 158; blue, 23 }  ,fill opacity=1 ][line width=2.25]  (117.5,228) .. controls (117.5,216.13) and (127.13,206.5) .. (139,206.5) .. controls (150.87,206.5) and (160.5,216.13) .. (160.5,228) .. controls (160.5,239.87) and (150.87,249.5) .. (139,249.5) .. controls (127.13,249.5) and (117.5,239.87) .. (117.5,228) -- cycle ;
\draw  [fill={rgb, 255:red, 57; green, 158; blue, 23 }  ,fill opacity=1 ][line width=2.25]  (205.5,228) .. controls (205.5,216.13) and (215.13,206.5) .. (227,206.5) .. controls (238.87,206.5) and (248.5,216.13) .. (248.5,228) .. controls (248.5,239.87) and (238.87,249.5) .. (227,249.5) .. controls (215.13,249.5) and (205.5,239.87) .. (205.5,228) -- cycle ;
\draw  [fill={rgb, 255:red, 57; green, 158; blue, 23 }  ,fill opacity=1 ][line width=2.25]  (117.5,318) .. controls (117.5,306.13) and (127.13,296.5) .. (139,296.5) .. controls (150.87,296.5) and (160.5,306.13) .. (160.5,318) .. controls (160.5,329.87) and (150.87,339.5) .. (139,339.5) .. controls (127.13,339.5) and (117.5,329.87) .. (117.5,318) -- cycle ;
\draw  [fill={rgb, 255:red, 57; green, 158; blue, 23 }  ,fill opacity=1 ][line width=2.25]  (205.5,318) .. controls (205.5,306.13) and (215.13,296.5) .. (227,296.5) .. controls (238.87,296.5) and (248.5,306.13) .. (248.5,318) .. controls (248.5,329.87) and (238.87,339.5) .. (227,339.5) .. controls (215.13,339.5) and (205.5,329.87) .. (205.5,318) -- cycle ;
\draw [line width=0.75]    (73.5,228) -- (115.5,228) ;
\draw [shift={(117.5,228)}, rotate = 180] [color={rgb, 255:red, 0; green, 0; blue, 0 }  ][line width=0.75]    (17.49,-5.26) .. controls (11.12,-2.23) and (5.29,-0.48) .. (0,0) .. controls (5.29,0.48) and (11.12,2.23) .. (17.49,5.26)   ;
\draw [line width=0.75]    (160.5,318) -- (203.5,318) ;
\draw [shift={(205.5,318)}, rotate = 180] [color={rgb, 255:red, 0; green, 0; blue, 0 }  ][line width=0.75]    (17.49,-5.26) .. controls (11.12,-2.23) and (5.29,-0.48) .. (0,0) .. controls (5.29,0.48) and (11.12,2.23) .. (17.49,5.26)   ;
\draw [line width=0.75]    (52,249.5) .. controls (52.5,309.7) and (46.06,319.4) .. (116.43,318.02) ;
\draw [shift={(117.5,318)}, rotate = 178.8] [color={rgb, 255:red, 0; green, 0; blue, 0 }  ][line width=0.75]    (17.49,-5.26) .. controls (11.12,-2.23) and (5.29,-0.48) .. (0,0) .. controls (5.29,0.48) and (11.12,2.23) .. (17.49,5.26)   ;
\draw [line width=0.75]    (140.83,250.32) .. controls (170.15,263.26) and (174.25,257.21) .. (211,243) ;
\draw [shift={(139,249.5)}, rotate = 24.3] [color={rgb, 255:red, 0; green, 0; blue, 0 }  ][line width=0.75]    (17.49,-5.26) .. controls (11.12,-2.23) and (5.29,-0.48) .. (0,0) .. controls (5.29,0.48) and (11.12,2.23) .. (17.49,5.26)   ;
\draw [line width=0.75]    (139,206.5) .. controls (160.07,193.27) and (181.14,200.21) .. (207.39,212.73) ;
\draw [shift={(209,213.5)}, rotate = 205.71] [color={rgb, 255:red, 0; green, 0; blue, 0 }  ][line width=0.75]    (17.49,-5.26) .. controls (11.12,-2.23) and (5.29,-0.48) .. (0,0) .. controls (5.29,0.48) and (11.12,2.23) .. (17.49,5.26)   ;
\draw [line width=0.75]    (240,301) .. controls (250.29,276.99) and (249.06,274.59) .. (238.65,250.03) ;
\draw [shift={(238,248.5)}, rotate = 67.07] [color={rgb, 255:red, 0; green, 0; blue, 0 }  ][line width=0.75]    (17.49,-5.26) .. controls (11.12,-2.23) and (5.29,-0.48) .. (0,0) .. controls (5.29,0.48) and (11.12,2.23) .. (17.49,5.26)   ;
\draw  [fill={rgb, 255:red, 57; green, 158; blue, 23 }  ,fill opacity=1 ][line width=2.25]  (31.5,393.25) .. controls (31.5,381.38) and (41.13,371.75) .. (53,371.75) .. controls (64.87,371.75) and (74.5,381.38) .. (74.5,393.25) .. controls (74.5,405.12) and (64.87,414.75) .. (53,414.75) .. controls (41.13,414.75) and (31.5,405.12) .. (31.5,393.25) -- cycle ;
\draw  [fill={rgb, 255:red, 57; green, 158; blue, 23 }  ,fill opacity=1 ][line width=2.25]  (118.5,393.25) .. controls (118.5,381.38) and (128.13,371.75) .. (140,371.75) .. controls (151.87,371.75) and (161.5,381.38) .. (161.5,393.25) .. controls (161.5,405.12) and (151.87,414.75) .. (140,414.75) .. controls (128.13,414.75) and (118.5,405.12) .. (118.5,393.25) -- cycle ;
\draw  [fill={rgb, 255:red, 57; green, 158; blue, 23 }  ,fill opacity=1 ][line width=2.25]  (206.5,393.25) .. controls (206.5,381.38) and (216.13,371.75) .. (228,371.75) .. controls (239.87,371.75) and (249.5,381.38) .. (249.5,393.25) .. controls (249.5,405.12) and (239.87,414.75) .. (228,414.75) .. controls (216.13,414.75) and (206.5,405.12) .. (206.5,393.25) -- cycle ;
\draw  [fill={rgb, 255:red, 57; green, 158; blue, 23 }  ,fill opacity=1 ][line width=2.25]  (118.5,483.25) .. controls (118.5,471.38) and (128.13,461.75) .. (140,461.75) .. controls (151.87,461.75) and (161.5,471.38) .. (161.5,483.25) .. controls (161.5,495.12) and (151.87,504.75) .. (140,504.75) .. controls (128.13,504.75) and (118.5,495.12) .. (118.5,483.25) -- cycle ;
\draw  [fill={rgb, 255:red, 57; green, 158; blue, 23 }  ,fill opacity=1 ][line width=2.25]  (206.5,483.25) .. controls (206.5,471.38) and (216.13,461.75) .. (228,461.75) .. controls (239.87,461.75) and (249.5,471.38) .. (249.5,483.25) .. controls (249.5,495.12) and (239.87,504.75) .. (228,504.75) .. controls (216.13,504.75) and (206.5,495.12) .. (206.5,483.25) -- cycle ;
\draw [line width=0.75]    (74.5,393.25) -- (116.5,393.25) ;
\draw [shift={(118.5,393.25)}, rotate = 180] [color={rgb, 255:red, 0; green, 0; blue, 0 }  ][line width=0.75]    (17.49,-5.26) .. controls (11.12,-2.23) and (5.29,-0.48) .. (0,0) .. controls (5.29,0.48) and (11.12,2.23) .. (17.49,5.26)   ;
\draw [line width=0.75]    (161.5,483.25) -- (204.5,483.25) ;
\draw [shift={(206.5,483.25)}, rotate = 180] [color={rgb, 255:red, 0; green, 0; blue, 0 }  ][line width=0.75]    (17.49,-5.26) .. controls (11.12,-2.23) and (5.29,-0.48) .. (0,0) .. controls (5.29,0.48) and (11.12,2.23) .. (17.49,5.26)   ;
\draw [line width=0.75]    (53,414.75) .. controls (53.5,474.95) and (47.06,484.65) .. (117.43,483.27) ;
\draw [shift={(118.5,483.25)}, rotate = 178.8] [color={rgb, 255:red, 0; green, 0; blue, 0 }  ][line width=0.75]    (17.49,-5.26) .. controls (11.12,-2.23) and (5.29,-0.48) .. (0,0) .. controls (5.29,0.48) and (11.12,2.23) .. (17.49,5.26)   ;
\draw [line width=0.75]    (141.83,415.57) .. controls (171.15,428.51) and (175.25,422.46) .. (212,408.25) ;
\draw [shift={(140,414.75)}, rotate = 24.3] [color={rgb, 255:red, 0; green, 0; blue, 0 }  ][line width=0.75]    (17.49,-5.26) .. controls (11.12,-2.23) and (5.29,-0.48) .. (0,0) .. controls (5.29,0.48) and (11.12,2.23) .. (17.49,5.26)   ;
\draw [line width=0.75]    (140,371.75) .. controls (161.07,358.52) and (182.14,365.46) .. (208.39,377.98) ;
\draw [shift={(210,378.75)}, rotate = 205.71] [color={rgb, 255:red, 0; green, 0; blue, 0 }  ][line width=0.75]    (17.49,-5.26) .. controls (11.12,-2.23) and (5.29,-0.48) .. (0,0) .. controls (5.29,0.48) and (11.12,2.23) .. (17.49,5.26)   ;
\draw [line width=0.75]    (241,466.25) .. controls (251.29,442.24) and (250.06,439.84) .. (239.65,415.28) ;
\draw [shift={(239,413.75)}, rotate = 67.07] [color={rgb, 255:red, 0; green, 0; blue, 0 }  ][line width=0.75]    (17.49,-5.26) .. controls (11.12,-2.23) and (5.29,-0.48) .. (0,0) .. controls (5.29,0.48) and (11.12,2.23) .. (17.49,5.26)   ;
\draw  [color={rgb, 255:red, 208; green, 2; blue, 27 }  ,draw opacity=1 ][dash pattern={on 5.63pt off 4.5pt}][line width=1.5]  (27,197.5) -- (179.33,197.5) -- (179.33,262.5) -- (27,262.5) -- cycle ;
\draw  [color={rgb, 255:red, 0; green, 0; blue, 0 }  ,draw opacity=1 ][dash pattern={on 5.63pt off 4.5pt}][line width=1.5]  (27,362.17) -- (179.33,362.17) -- (179.33,427.17) -- (27,427.17) -- cycle ;
\draw  [color={rgb, 255:red, 208; green, 2; blue, 27 }  ,draw opacity=1 ][dash pattern={on 5.63pt off 4.5pt}][line width=1.5]  (190.5,362.17) -- (252,362.17) -- (252,508) -- (190.5,508) -- cycle ;
\draw  [color={rgb, 255:red, 208; green, 2; blue, 27 }  ,draw opacity=1 ][dash pattern={on 5.63pt off 4.5pt}][line width=1.5]  (92,457.33) -- (178.67,457.33) -- (178.67,509.33) -- (92,509.33) -- cycle ;
\draw  [line width=1.5]  (23.33,37.03) .. controls (23.33,34.03) and (25.76,31.6) .. (28.76,31.6) -- (266.57,31.6) .. controls (269.57,31.6) and (272,34.03) .. (272,37.03) -- (272,521.9) .. controls (272,524.9) and (269.57,527.33) .. (266.57,527.33) -- (28.76,527.33) .. controls (25.76,527.33) and (23.33,524.9) .. (23.33,521.9) -- cycle ;
\draw [line width=1.5]    (22.67,180.33) -- (272,180.33) ;
\draw [line width=1.5]    (22.67,345.33) -- (272,345.33) ;

\draw (42,53.25) node [anchor=north west][inner sep=0.75pt]  [font=\LARGE,color={rgb, 255:red, 255; green, 255; blue, 255 }  ,opacity=1 ] [align=left] {\textbf{1}};
\draw (129.5,53.25) node [anchor=north west][inner sep=0.75pt]  [font=\LARGE] [align=left] {\textbf{\textcolor[rgb]{1,1,1}{2}}};
\draw (218.5,53.25) node [anchor=north west][inner sep=0.75pt]  [font=\LARGE] [align=left] {\textbf{\textcolor[rgb]{1,1,1}{3}}};
\draw (217.5,144.75) node [anchor=north west][inner sep=0.75pt]  [font=\LARGE] [align=left] {\textbf{\textcolor[rgb]{1,1,1}{4}}};
\draw (130,144.75) node [anchor=north west][inner sep=0.75pt]  [font=\LARGE] [align=left] {\textbf{\textcolor[rgb]{1,1,1}{5}}};
\draw (28,6.5) node [anchor=north west][inner sep=0.75pt]  [font=\LARGE] [align=left] {{\LARGE \textbf{Example}}};
\draw (253,158.5) node [anchor=north west][inner sep=0.75pt]  [rotate=-270] [align=left] {\textbf{{\Large Iteration 1}}};
\draw (42.5,216.75) node [anchor=north west][inner sep=0.75pt]  [font=\LARGE,color={rgb, 255:red, 255; green, 255; blue, 255 }  ,opacity=1 ] [align=left] {\textbf{1}};
\draw (130,216.75) node [anchor=north west][inner sep=0.75pt]  [font=\LARGE] [align=left] {\textbf{\textcolor[rgb]{1,1,1}{2}}};
\draw (219,216.75) node [anchor=north west][inner sep=0.75pt]  [font=\LARGE] [align=left] {\textbf{\textcolor[rgb]{1,1,1}{3}}};
\draw (219,308.25) node [anchor=north west][inner sep=0.75pt]  [font=\LARGE] [align=left] {\textbf{\textcolor[rgb]{1,1,1}{4}}};
\draw (130.5,308.25) node [anchor=north west][inner sep=0.75pt]  [font=\LARGE] [align=left] {\textbf{\textcolor[rgb]{1,1,1}{5}}};
\draw (253.5,322) node [anchor=north west][inner sep=0.75pt]  [rotate=-270] [align=left] {\textbf{{\Large Iteration 2}}};
\draw (44.5,382) node [anchor=north west][inner sep=0.75pt]  [font=\LARGE,color={rgb, 255:red, 255; green, 255; blue, 255 }  ,opacity=1 ] [align=left] {\textbf{1}};
\draw (132,382) node [anchor=north west][inner sep=0.75pt]  [font=\LARGE] [align=left] {\textbf{\textcolor[rgb]{1,1,1}{2}}};
\draw (221,382) node [anchor=north west][inner sep=0.75pt]  [font=\LARGE] [align=left] {\textbf{\textcolor[rgb]{1,1,1}{3}}};
\draw (219,473.5) node [anchor=north west][inner sep=0.75pt]  [font=\LARGE] [align=left] {\textbf{\textcolor[rgb]{1,1,1}{4}}};
\draw (130.5,473.5) node [anchor=north west][inner sep=0.75pt]  [font=\LARGE] [align=left] {\textbf{\textcolor[rgb]{1,1,1}{5}}};
\draw (254.5,487.25) node [anchor=north west][inner sep=0.75pt]  [rotate=-270] [align=left] {\textbf{{\Large Iteration 3}}};
\draw (23.67,182.08) node [anchor=north west][inner sep=0.75pt]  [font=\large,color={rgb, 255:red, 208; green, 2; blue, 27 }  ,opacity=1 ] [align=left] {\textbf{Server 1}};
\draw (187.67,347.42) node [anchor=north west][inner sep=0.75pt]  [font=\large,color={rgb, 255:red, 208; green, 2; blue, 27 }  ,opacity=1 ] [align=left] {\textbf{Server 2}};
\draw (92,510.33) node [anchor=north west][inner sep=0.75pt]  [font=\large,color={rgb, 255:red, 208; green, 2; blue, 27 }  ,opacity=1 ] [align=left] {\textbf{Server 3}};
\draw (22.67,345.33) node [anchor=north west][inner sep=0.75pt]  [font=\large,color={rgb, 255:red, 0; green, 0; blue, 0 }  ,opacity=1 ] [align=left] {\textbf{Server 1}};

\end{tikzpicture}};
  };
  \end{tikzpicture}
}

There may be situations where, at the end of an iteration of the outer while-loop, one or more flow constraints~\eqref{COM.c4} are violated -- this can happen when graph $(\mathcal{M}^{(t)}, \mathbf{V}^{(t)})$ is densely connected, and link capacities vary greatly. Flow constraint violations are filtered out at line~23.
For each iteration of the mapping task, the inner while-loop iterates until $\mathbf{K}$ is empty that is no more than $2|\mathcal{M}^{(t)}| - 1$ times -- equal to the maximum number of times the GP ($|\mathcal{M}^{(t)}| - 1$) and BFF ($|\mathcal{{M}}^{(t)}|$) procedure may run.
\begin{remark}
Adopting the auxiliary counters $\omega_i$ in Procedure~\ref{alg:MMP} ensures that each MS instance is mapped on a server only once; hence, constraint~\eqref{COM.c1} is met.
The server selection method adopted by the BFF procedure ensures that constraint~\eqref{COM.c2} is met.
The BFF procedure also ensures both constraints~\eqref{COM.c3} and~\eqref{COM.c4} are fulfilled. Hence, Procedure~\ref{alg:MMP} returns feasible solutions to the CO problem.
\end{remark}

\vspace{-5mm}\section{Analytical Results}\label{sec:evaluation}\vspace{-2mm}
In this section, we will refer to an equivalent formulation of the CO problem where the terms involving products of pairs of binary optimization variables have been linearized by adopting the transformation defined in~\cite[Eq.~(3)-(7)]{math10020283}. In doing so, the resulting equivalent optimization problem is linear.

\begin{figure}[t]
\vspace{-6.5mm}
\centering
\hspace*{-1.5mm}\subfloat[Homogeneous Case]{\label{fig.1.1}
    \includegraphics[width=0.5\columnwidth]{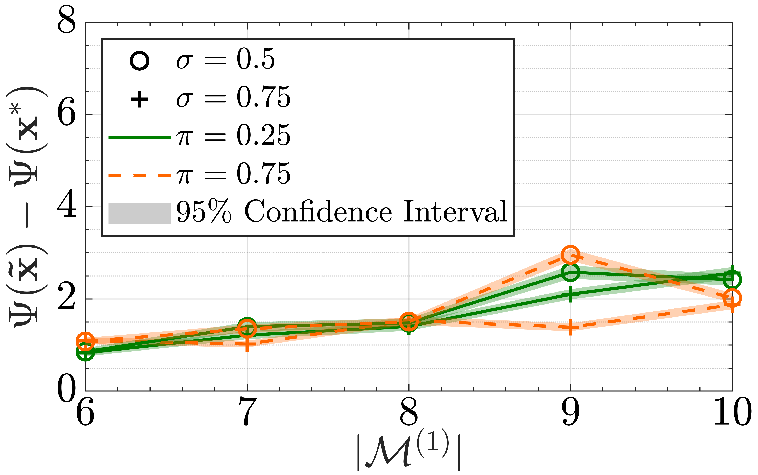}
}
\hspace*{-1.8mm}\subfloat[Non-Homogeneous Case]{\label{fig.1.2}
    \includegraphics[width=0.5\columnwidth]{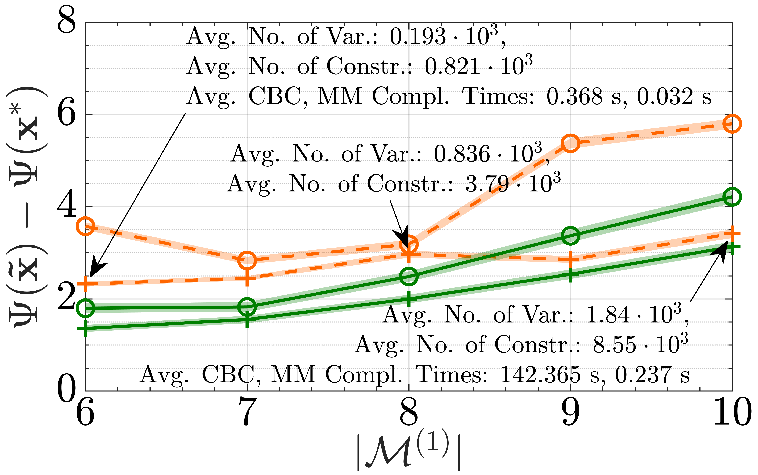}
}
\vspace{-1mm}
\caption{Average cost difference as a function of the total number of MS instances to be mapped onto the available servers. The average CBC and MM procedure completion times have been measured on a workstation equipped with a CPU AMD 5995WX. Legend of both figures is reported in Fig.~\ref{fig.1.1}.}
\label{fig.1}\vspace{-3mm}
\end{figure}

Let $\Psi(\mathbf{x}^*)$ and $\Psi(\mathbf{\tilde{x}})$ be the values of the objective function~\eqref{COM.of1} when the solution ($\mathbf{x}^*$) to the CO problem is calculated via the Coin-or Branch and Cut (CBC) solver~\cite{john_forrest_2023_7843975} and when the solution ($\mathbf{\tilde{x}}$) is obtained via the proposed MM procedure, respectively. Since $\Psi(\tilde{\mathbf{x}}) \geq \Psi(\mathbf{x}^*)$, Fig.~\ref{fig.1} establishes how much a solution $\mathbf{\tilde{x}}$ deviates from $\mathbf{x}^*$ by showing the \emph{cost difference} $\Psi(\mathbf{\tilde{x}}) - \Psi(\mathbf{x}^*)$ as a function of the total number of MS instances to be mapped onto the available servers.

Fig.~\ref{fig.1} refers to scenarios where a single MS-based procedure ($t = 1$) is considered; the same scenarios would apply to an NF-based architecture. To effectively investigate the performance of the proposed MM procedure, we considered a variable number of MS instances $|\mathcal{M}^{(1)}| = \{6, \ldots, 10\}$, where matrix $\mathbf{V}^{(1)}$ has been generated uniformly at random such that the probability of an instance to communicate with another directly is $\pi = \{0.25, 0.75\}$ -- for each of the considered number of MS instances $|\mathcal{M}^{(1)}|$ we employed $500$ Monte Carlo iterations. The (normalized) CPU and memory footprints ($\mathcal{F}^{(1)}$) of each instance are set equal to $1$, while the processing time components are set according to the following relations:  $a^{(1)}_{i,j} = \SI{1}{\milli\second}$, and $c^{(1)}_i = \SI{0.5}{\milli\second}$, for $i,j \in \mathcal{M}^{(1)}$. Finally, the maximum number $\ell_i^{(1)}$ of concurrent requests that each MS supports is set equal to $1$, for $i \in \mathcal{M}^{(1)}$.

Fig.~\ref{fig.1.1} shows results for deployments consisting of $|\mathcal{S}| = \lceil\sigma |\mathcal{M}^{(1)}|\rceil$ servers, for $\sigma = \{0.5, 0.75\}$. The (normalized) CPU resource capacity $r^{(\textrm{CPU})}_i$ of each server is set equal to $\lceil(\sum_{i \in \mathcal{M}^{(1)}} f^{(1,\textrm{CPU})}_i)/|\mathcal{S}|\rceil$ (\emph{homogeneous case}). Similar considerations apply to the memory resource capacity. Any server can directly communicate with any other server (full connectivity assumption), and each outgoing network link has a capacity equal to the total flow generated by the maximum number of MS instances that can be run on a server.

In the considered scenarios, Fig.~\ref{fig.1.1} shows that the average cost difference (averaged across all the randomly generated instances of matrix $\mathbf{V}^{(1)}$) tends to increase as the total number of considered MS instances increases. The larger the number of MS instances to be mapped onto servers, the harder the CO problem becomes both in terms of the number of optimization variables and the total number of constraints. Regardless, in the scenarios characterized by the largest number of servers ($\sigma = 0.75$) and the highest probability of an MS instance directly communicating with another one ($\pi = 0.75$), the average cost difference increases from $1.01$ (for $|\mathcal{M}^{(1)}| = 6$) to $1.9$ (for $|M^{(1)}| = 10$). Overall, different values of $\sigma$ and $\pi$ appear not to impact the average cost difference significantly.

Fig.~\ref{fig.1.2} considers scenarios equivalent to those associated with Fig.~\ref{fig.1.1} except that the resource capacity of the servers is no longer homogeneous (\emph{non-homogeneous case}). In this case, the CPU resource capacity of $25\%$ of the available servers is three times smaller ($r^{(\textrm{CPU})}_i = \lceil(\sum_{i \in \mathcal{M}^{(1)}} f^{(1,\textrm{CPU})}_i)/(3|\mathcal{S}|)\rceil$) than the resource capacities of the remaining $75\%$ ($r^{(\textrm{CPU})}_i = \lceil(2\sum_{i \in M^{(1)}} f^{(1,\textrm{CPU})}_i)/(3|S|)\rceil$), similar considerations apply for the memory resource capacity.  This change in CPU and memory resource capacities is used to illustrate results for a network that includes lower capability edge servers, whereas Fig.~\ref{fig.1.1} can be considered to be a deployment model of only centralized or cloud-based servers.

Similarly to Fig.~\ref{fig.1.1}, Fig.~\ref{fig.1.2} shows that the average cost difference tends to increase with $|\mathcal{M}^{(1)}|$.
As expected, the average cost difference is larger than the corresponding cases shown in Fig.~\ref{fig.1.1} -- this is ultimately caused by the non-homogeneity of server resource capacity that makes the bin-packing challenging. In particular, this applies to scenarios where MS instances are more densely interconnected ($\pi = 0.75$) than in those scenarios where MS instances are less interconnected ($\pi = 0.5$). Yet, the average cost difference appears to be greater in those scenarios associated with the smallest number of servers ($\sigma = 0.5$). This can be intuitively explained by the fact that the smaller the value of $\sigma$, the more the MM procedure will favor the reduction in the overall servers needed to run a given set of MS instances over the inter-server traffic. Regardless, it is also worth noting that the largest average cost difference is smaller than $5.78$ (for $|\mathcal{M}^{(1)} = 10|$, $\sigma = 0.5$, and $\pi = 0.75$).

For each of the considered values of $|\mathcal{M}^{(1)}|$, Fig.~\ref{fig.1.2} also highlights the largest average number of optimization variables and constraints forming the instances of the linearized version of the CO problem, which are associated with scenarios where $\sigma = 0.75$ and $\pi = 0.75$. When $|\mathcal{M}^{(1)}|$ increases from $6$ to $10$, we observe that the average number of optimization variables (constraints) increases by a factor of $9.55$ ($10.41$). At the same time, the average time the CBC solver (MM procedure) needs to solve the problem increases from $\SI{0.37}{\second}$ to $\SI{142.36}{\second}$ (from $\SI{0.032}{\second}$ to $\SI{0.24}{\second}$) -- thus, showing how impractical it would be to attempt to solve the CO problem directly and reinforcing the need of formulating a heuristic strategy.

\begin{figure}[t]
\vspace{-7mm}
\centering
\hspace*{-1.5mm}\subfloat[$\Hat{I}$]{\label{fig.2.3}
    \includegraphics[width=0.5\columnwidth]{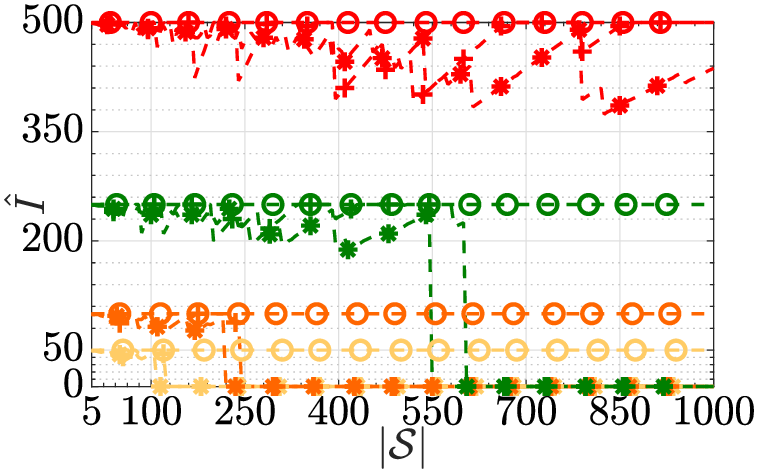}
}
\hspace*{-1.8mm}\subfloat[$\Psi(\mathbf{\tilde{x}})$]{\label{fig.2.4}
    \includegraphics[width=0.5\columnwidth]{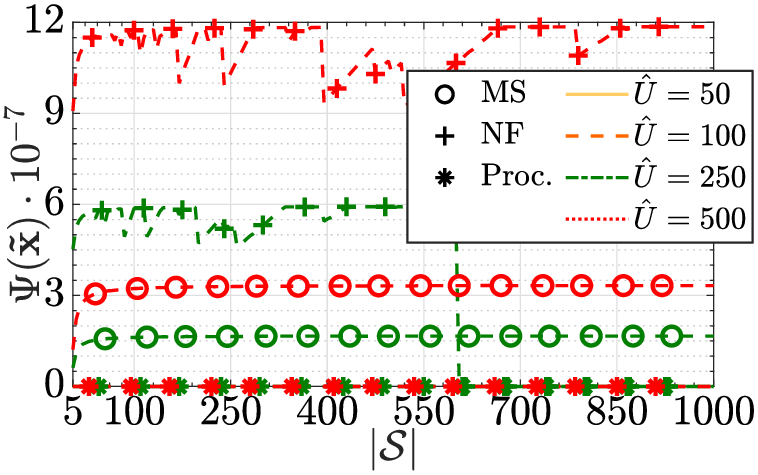}
}
\vspace{-1.5mm}
\caption{Maximum number of instances and cost as a function of $|\mathcal{S}|$, for multiple requested loads, in the case of MS-, NF- and procedure-based architectures. Legend of all figures is reported in Fig.~\ref{fig.2.4}.}\vspace{-2mm}
\label{fig.2}
\end{figure}

We now compare the performance of MS-based, NF-based, and procedure-based core network architectures as a function of the number of available servers in non-homogeneous scenarios. We considered an ideal core network model offering three CP procedures: UE Registration, UE Deregistration, and PDU Session Modification~\cite{3gpp.23.502}.
In the case of an MS-based architecture, we refer to Remark~\ref{rem.implem}, thus, leading to a total number of MSs equal to $62$ ($34$ for UE Registration, $15$ for UE Deregistration, and $13$ for PDU Session Modification).
For each MS, we assumed the same resource footprint model considered in Fig.~\ref{fig.1}. We modeled an NF-based architecture by considering the 5GC NFs, where the CPU footprint of each NF is set equal to the overall footprint of each MS associated with a service-operation pair pertaining to the same NF; similar considerations apply to memory footprint and output flow (expressed in terms of number of PDUs per second).
The procedure-based architecture consists of three types of entities (one per procedure); each entity's CPU and memory footprint is set equal to the overall footprint of all the MSs pertaining to the corresponding procedure. The maximum number of concurrent requests that each MS, NF, or procedural entity can support is set equal to $1$. The allocation of MS, NF, and procedural entity instances onto servers is carried out employing the proposed MM heuristic.

Fig.~\ref{fig.2.3} shows the actual number $\hat{I}$ of concurrent user requests each procedure can support as a function of $|\mathcal{S}|$, when $\hat{U}^{(t)} = \hat{U} = {50, 100, 250, 500}$, for $t = 1, \ldots, 3$. Although the resource capacity of each server decreases as $|\mathcal{S}|$ increases, the MS-based architecture can always support the maximum number of concurrent user requests; this is made possible by the reduced footprint of each MS instance. Conversely, since some NF instances determine a greater resource footprint, finding servers with sufficient capacity to allocate them gets harder and harder. This determines a reduction and sharp fall to zero of $\hat{I}$, which occurs when the server resource capacity is no longer enough to run any instances of the NF associated with the largest footprint. Similar considerations apply to the procedure-based architecture; when $\hat{U} = 250$, for the NF- and procedure-based architectures, the value of $\hat{I}$ drops to zero for $|\mathcal{S}|$ equal to $605$ and $550$, respectively. In the NF- and procedure-based architectures, we also observe that the oscillating behavior in the value of $\hat{I}$ occurs because NF and procedure entity instances require a non-negligible resource footprint that may not be available in full as $|S|$ increases (server capacity reduces).

Fig.~\ref{fig.2.4} compares the cost (see Eq.~\eqref{COM.of1}) associated with each architecture as a function of $|\mathcal{S}|$ (for $\hat{U} = {250, 500}$). In the procedure-based case, since each procedural entity delivers an entire CP procedure, it must be allocated on a single server, and thus, inter-server messaging cost is always equal to zero. Conversely, the cost associated with an NF-based architecture is up to twice as much as the cost associated with an MS-based architecture since each NF instance is expected to carry out a set of functionalities comprised of multiple MS instances, and NFs are often forced onto separate servers. Each NF instance's maximum output flow is higher than any MS instance's output flow, which translates into an overall higher cost when different NF instances are mapped onto different servers. We also observe that the cost associated with an MS-based architecture eventually plateaus. This directly follows from the fact that the rate at which the average number of MS instances mapped on each server decays becomes smaller, the larger $|\mathcal{S}|$ becomes -- the same considerations apply to the NF- and procedure-based cases (up until their cost drops to zero).

\begin{figure}[t]
    \vspace{-2mm}
    \centering
    \includegraphics[width=\columnwidth]{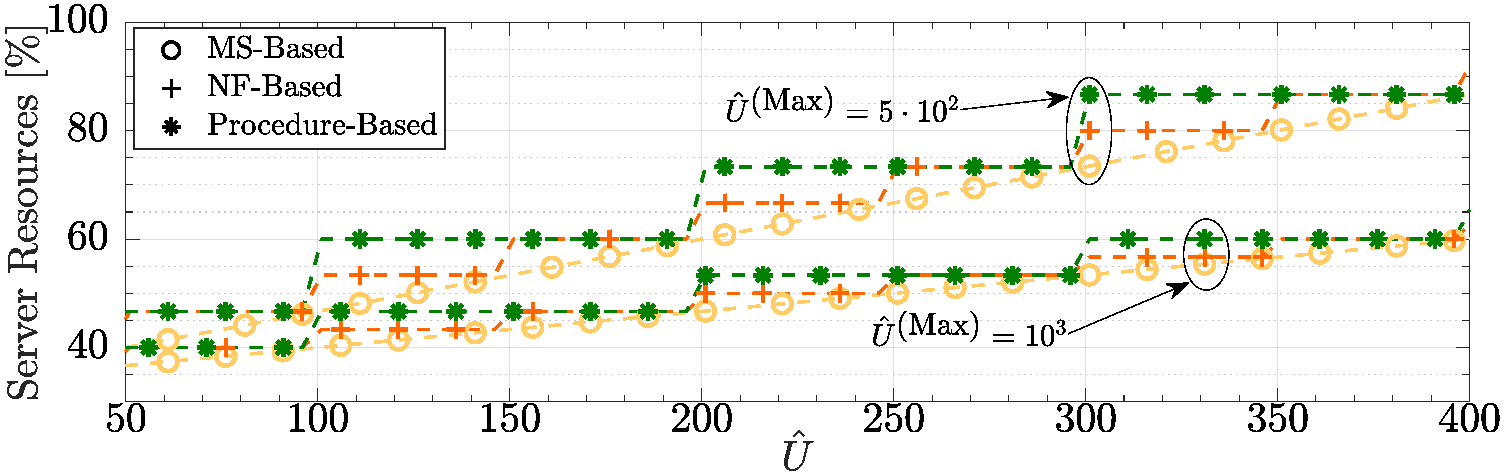}
    \vspace{-7.5mm}
    \caption{Allocated bare metal resources (CPU or memory) as a function of $\hat{U}$, for MS-based and multi-threaded NF- and procedure-based architectures.}\vspace{-2mm}
    \label{fig.3}
\end{figure}

Fig.~\ref{fig.3} considers a family of non-homogenous scenarios where the number of servers is fixed to five $|\mathcal{S}| = 5$. Besides, each NF and procedure entity instance can serve multiple requests simultaneously (\emph{multi-threaded implementation}), which we assumed to be $50$ and $100$ requests for NF- and procedure-base architectures, respectively. The CPU/memory footprint of each NF and procedure entity instance is identical to the one considered in Fig.~\ref{fig.2} multiplied for the maximum number of concurrent requests served. We regard $\hat{U}^{(\textrm{Max})}$ as the maximum number of concurrent user requests for each CP procedure in the case of an MS-based architecture if the overall bare metal capacities were collated in a single server. Fig.~\ref{fig.3} shows the percentage of allocated bare metal resources (CPU or memory) as a function of $\hat{U}$. Due to its high level of granularity, the MS-based architecture can accurately track the current requested load. Multi-threaded implementations of NF and procedure entity instances ensure an increased processing bandwidth at the cost of a raised resource footprint. Overall, for $\hat{U} = 300$ and $\hat{U}^{(\textrm{Max})} = 500$, an MS-based architecture requires $~70\%$ of the overall bare metal resources, an NF-based architecture $~80\%$, and over $86\%$ for a procedure-based architecture.

\vspace{-4mm}\section{Conclusions}\vspace*{-1mm}
\label{sec:conclusion}
We formulated a novel optimization problem to partition and allocate large-scale microservice graphs of CP procedures onto heterogeneous bare metal deployments while containing the total network traffic among servers. 
We also proposed an efficient heuristic strategy to solve the problem above. In the considered scenarios, solutions obtained with the proposed heuristic procedure only marginally deviated from solutions obtained with a branch-and-cut approach. Analytical results establish that an MS-based core network consistently supports the requested CP load in heterogeneous bare metal deployments while NF- and procedure-based cores performance quickly degrades (as the server resource capacity reduces). Our optimization framework provides an MS-server allocation solution, ensuring an overall reduction in the CP-related network traffic if compared to an NF-based core. Future research avenues involve the accurate profiling of MS's computational and network footprints and the extension of the optimization framework to make it possible to continuously re-optimize the MS instance allocation by accounting for user requests and bare metal resources that change over time.

\bibliographystyle{IEEEtran}
\vspace{-4mm}\bibliography{bibliography}\vspace*{-20mm}

\end{document}